\documentclass[superscriptaddress, 
10pt,onecolumn,oneside,a4paper,aps 
]{revtex4-1}
\usepackage{graphicx}
\usepackage{dcolumn}
\usepackage{bm}
\usepackage{subcaption}
\usepackage{float} 
\usepackage{natbib}
\usepackage{amsmath}
\usepackage{array, makecell}  
\usepackage{multirow}  
\usepackage{lipsum}
\usepackage{color,soul,xcolor} 
\soulregister\cite7
\usepackage{hyperref} 

\pdfoutput=1

\begin{document}


\title{Inequality in Death from Social Conflicts: A Gini \& Kolkata indices-based Study }

\author{Antika Sinha}
\email{antikasinha@gmail.com}
 \affiliation{Department of Computer Science, Asutosh College, Kolkata-700026, India}
\author{Bikas K. Chakrabarti}
 \email{bikask.chakrabarti@saha.ac.in}
\affiliation{Saha Institute of Nuclear Physics, Kolkata-700064, India}
\affiliation{S.N.Bose National Centre for Basic Sciences, Kolkata-700106, India}
\affiliation{Economic Research Unit, Indian Statistical Institute, Kolkata-700108, India}

\vskip 1.5 cm
\begin{abstract}

\noindent
Human deaths caused by individual man-made conflicts (e.g., wars, armed-conflicts, terrorist-attacks etc.) occur unequally across the events (conflicts) and such inequality (in deaths) have been studied here using Lorenz curve and values of the inequality indices Gini ($g$) and Kolkata ($k$) have been estimated from it. The data are taken from various well-known databases maintained by some Universities and Peace Research Institutes. The inequality measures for man-made conflicts are found to have very high values ($g$ = $0.82 ~\pm~ $0.02, $k$ = $0.84~ \pm~ $0.02), which is rarely seen in economic (income or wealth) inequality measures across the world ($g \leq 0.4$, $k \leq 0.6$; presumably because of various welfare measures). We also investigated the inequalities in human deaths from natural disasters (like earthquakes, floods, etc.).  Interestingly, we observe that the social inequality measures ($g$ and $k$ values) from man-made conflicts compare well with those of academic centers (inequality in citations; found in earlier studies) of different institutions of the world, while those for natural disasters can be even higher. We discuss about the `similarity classes' of social inequality (similar higher values of $g$ and $k$ indices) for man-made competitive societies like academic institutions and man-made social conflicts, and connect our observations with that of the growing recent trend of economic inequality across the world (with rapid disappearance of welfare strategies).

\end{abstract}

\maketitle

\section{Introduction}
\noindent
While studying social systems, we come across various stabilized inequalities, as results of inherent social dynamics. Sociophysics \cite{sen2014sociophysics, castellano2009statistical, galam2012sociophysics}     
attempts to study such inequalities and capture them in various models of social dynamics. Though not inevitable, death to mortal life in natural or man-made disasters seems to be a significant possibility in our life. Since ancient times, both of man-made conflicts as well as natural disasters have played their role in shaping the structure of human society. Sometimes of course the distinction between man-made disasters and natural disasters are not very clear: Malthus suggested \cite{malthus1888essay} that (in absence of any birth control) the population grows with time in geometric progression (exponentially), while food-grain supply (cultivable land, the basic ingredient for food-grain production) grows in algebraic progression (linear in time), suggesting that famines or wars will occur almost periodically. This and similar arguments suggest that even the events like wars, armed-conflicts or in some cases the famines, may be natural and not directly man-made. We will, however,  not consider such arguments and distinguish them using commonly used criteria (as given in various datasets studied here).

We study here the inequality measures (from the corresponding human death distributions) for all kinds of social conflicts (wars, violent crimes, etc.) and also the same for natural disasters (earthquakes, floods, etc.). The detailed nature of the death count distributions (fat tails in particular) in such conflicts and disasters have already been studied earlier \cite{chatterjee2017fat}. We utilized the data publicly available from various established data-sources maintained by various universities and Peace Research centers. We find, the extent of inequality (using the specific measures defined later) is much higher in social conflicts when compared with those observed for the economic inequalities (for income, wealth) in earlier decades. We find that the `similarity classes' of social inequality (similar values of $g$ and $k$ indices, defined in the next section) for man-made competitive societies like academic institutions and for social conflicts are the same. We connect our observations about extreme inequalities in various social institutions and in social conflicts with that of the growing recent trend of economic inequality across the world (with rapid disappearance of welfare strategies).

\section{Inequality indices}
Generally for measuring (income or wealth) inequality, Gini index ($g$) \cite{gini1921measurement} is used by the economists who constructs first the Lorenz curve (see Fig. 1), which gives the cumulative fraction of the wealth against the fraction of population (possessing that wealth), when they are arranged in increasing order of income or wealth. The equality line corresponds to the case when the people have equal income or wealth each.  We have used the equivalent Lorenz curve (see, Fig. 1), where cumulative number of deaths (instead of wealth) are plotted against the fraction of events/conflicts (instead of population), when arranged in increasing order of size of the events or conflicts. The area between the equality line and Lorenz curve  (when normalised) gives the Gini index ($g$). We will also consider the Kolkata index $k$ \cite{ghosh2014inequality,chatterjee2015social,chatterjee2017socio} here, which is given by the value of the $x$-coordinate of the crossing point of the diagonal perpendicular to the equality line and the Lorenz curve (see Fig. 1). It may be noted that $g$ takes values within the range 0 (representing complete equality) and 1 (representing complete inequality), while the corresponding $k$ values range from 0.5 to 1, and the values of $2k-1$ ranges from 0 to 1. In view of the fact that ($1-k$) fraction of people, papers or wars do posses, capture or cause exactly $k$ fraction of wealth, citations or deaths respectively, we give the $k$ index values in various cases considered here. Also, $k$ index corresponds to a non-trivial fixed point of the complementary (non-linear) Lorenz function $\tilde{F}(n) \equiv 1-F(n)$ (see Fig. 1): $\tilde{F}(k) = k$. It may be mentioned in this context that in ~\cite{sreenivasan2010tricks}, a decomposition of $g$ in terms of a quantity equal in magnitude of $k$ was attempted. There are several other social inequality measuring indices  \cite{eliazar2010measuring,eliazar2012power,eliazar2014social,eliazar2015sociogeometry,
eliazar2016harnessing}, based on the Lorenz curve properties, which we will not consider here.

\begin{figure}[H]
\centering
\includegraphics[width=0.305\textwidth] {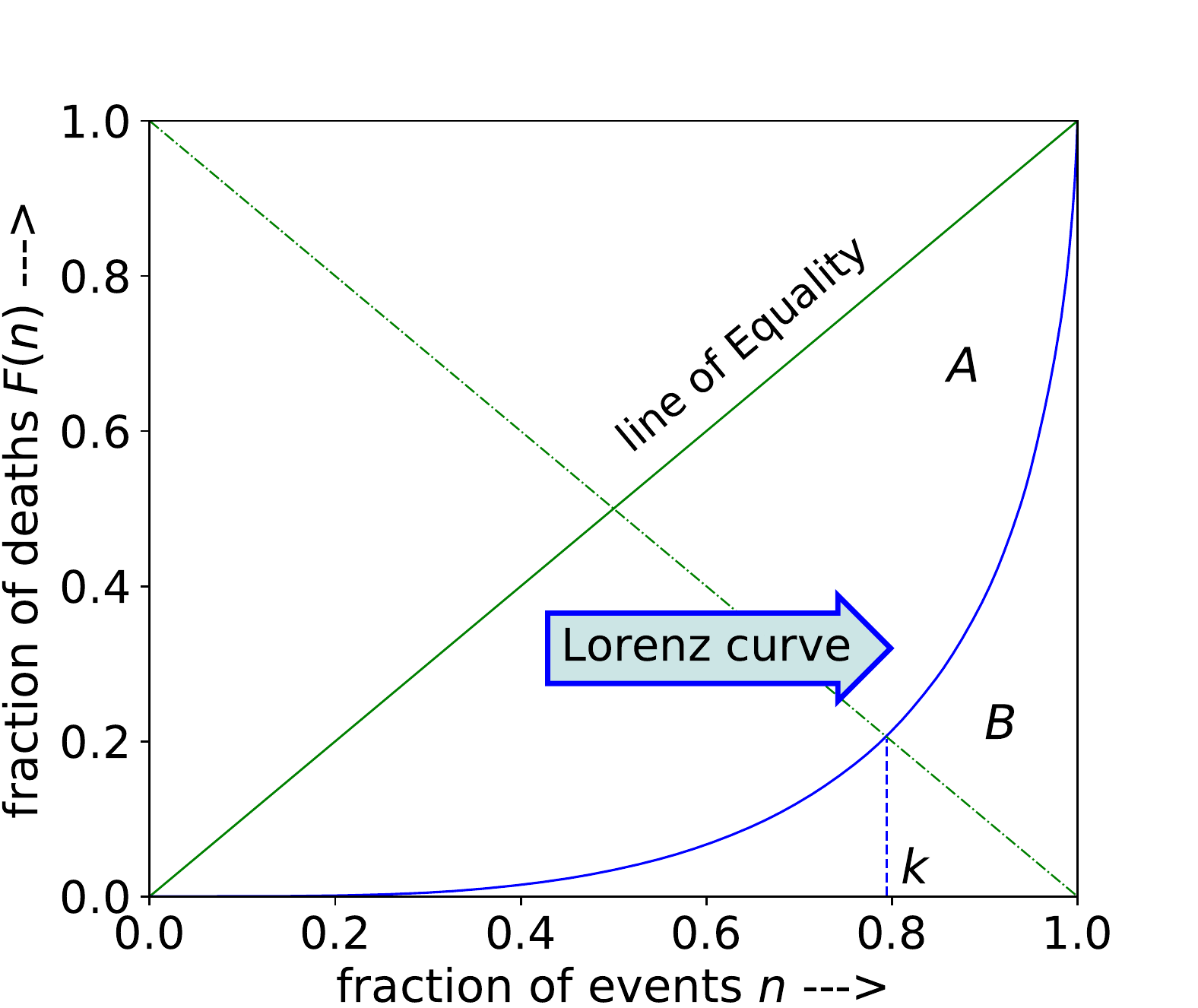}
\caption{Lorenz curve and inequality indices, Gini ($g$) and Kolkata ($k$). The cumulative fraction of deaths $F(n)$, when plotted against the fraction ($n$) of events or conflicts (arranged in increasing order of death counts) gives the Lorenz curve. Line of equality (solid diagonal) represents the case where each event/conflict causes identical number of deaths. The area $A$ (calculated in this work using Monte Carlo integration method) represents that between the Lorenz curve and the Equality line, and $B$ represents the area between the Lorenz curve and the right corner axes. Their normalised ratio gives the Gini index value: $\left(g={{A}  \over{A+B}}\right)$. When the other diagonal (perpendicular to the equality line) cuts the Lorenz curve at a point with x-coordinate value $k$, it means (1-$k$) fraction of events or conflicts have caused $k$ fraction of deaths; the Kolkata index value is given by $k$.}
\end{figure}

\section{DATASET DESCRIPTION}

We have studied here the statistics of human death in various man-made events or conflicts and natural disasters like earthquake etc. using the extensive and publicly available death count datasets from various universities and public research organisations for a long period of collections. Few data are found missing, and obviously get ignored in this study. The details of the used datasets are given below in Table I:

\begin{table}[ht]
  \begin{center}
    \caption{Dataset description of social damages (i.e. human deaths) used in our study}
    \label{tab:table1}
    \begin{tabular}{|c|c|c|c|c|c|c|c|c|} 
    \hline
    \textbf{\makecell{Disaster\\type}} & \textbf{Data source} & \textbf{\makecell{Data\\coverage}} & \textbf{Time period} & \textbf{\makecell{Dataset\\ length }} & \multicolumn{4}{ c|} {\textbf{human death count}}\\ \cline{6-9}   & & & & &  Min. & Max. & Avg. & Total  \\	
	\hline
       & \makecell{\textbf{war}\\CoW data (v.4.0) \cite{sarkees2016cow} } & worldwide & 1816-2007 & 538 & 0 & 1250000 & 28553 & 15361249 \\ \cline{2-9}

       & \makecell{\textbf{battle}\\UCDP data (v.18.1) \cite{uppsala2014ucdp}} & worldwide & 1989-2017 & 4695 & 25 & 500000  & 783 & 3674068 \\ \cline{2-9}
	   
 		man-made & \makecell{\textbf{armed-conflict}\\PRIO data(v.3.1) \cite{lacina2009battle}} & worldwide & 1946-2008 & 1186 & 13 & 497500 & 7175 & 8510077 \\ \cline{2-9}

     &  \makecell{\textbf{murder (violent-crime)}\\NCRB data \cite{ncrb}} & India & 1967-2016 & 1583 & 0 & 10776 & 902 & 1427276\\ \cline{2-9}

		&  \makecell{\textbf{terrorism}\\GTD data \cite{gtd}} & worldwide & 1970-2017 & 172109 & 0 & 1570 & 2-3  & 414537\\ \cline{2-9}
	   
      \hline
       & \makecell{\textbf{earthquake}\\NCEI-NOAA data \cite{ngdcearthquake}} & worldwide & 1000-2018(July) & 1321 & 1 & 830000 & 3312 & 4375172 \\ \cline{2-9}

		natural & \makecell{\textbf{flood}\\EMDAT data\cite{emdat}} & worldwide & 1900-2018(July) & 3540 & 1 & 3700000 & 1965 & 6957472 \\ \cline{2-9}

			 & \makecell{\textbf{tsunami}\\NCEI-NOAA data \cite{ngdctsunami}} & worldwide & 1000-2018(July) & 372 & 1 & 167540 & 1038  & 386275 \\ \cline{2-9}
        \hline
    \end{tabular}
  \end{center}
  \end{table}

\section{ANALYSIS AND ESTIMATES OF $g$ , $k$ VALUES}
\subsection{For man-made conflicts}

We used the human death data in various kinds of social conflicts, as recorded in various datasets e.g., CoW\cite{sarkees2016cow}, UCDP\cite{uppsala2014ucdp}, PRIO\cite{lacina2009battle}, NCRB\cite{ncrb} and GTD\cite{gtd}. The estimated Lorenz curves (shown in Figs. 2a, 2b, 2c) give the values of the Gini ($g$) and Kolkata ($k$) indices. These indices have quite high values (see Table II).

\begin{figure}[H]
\centering
\includegraphics[width=1.0\textwidth] {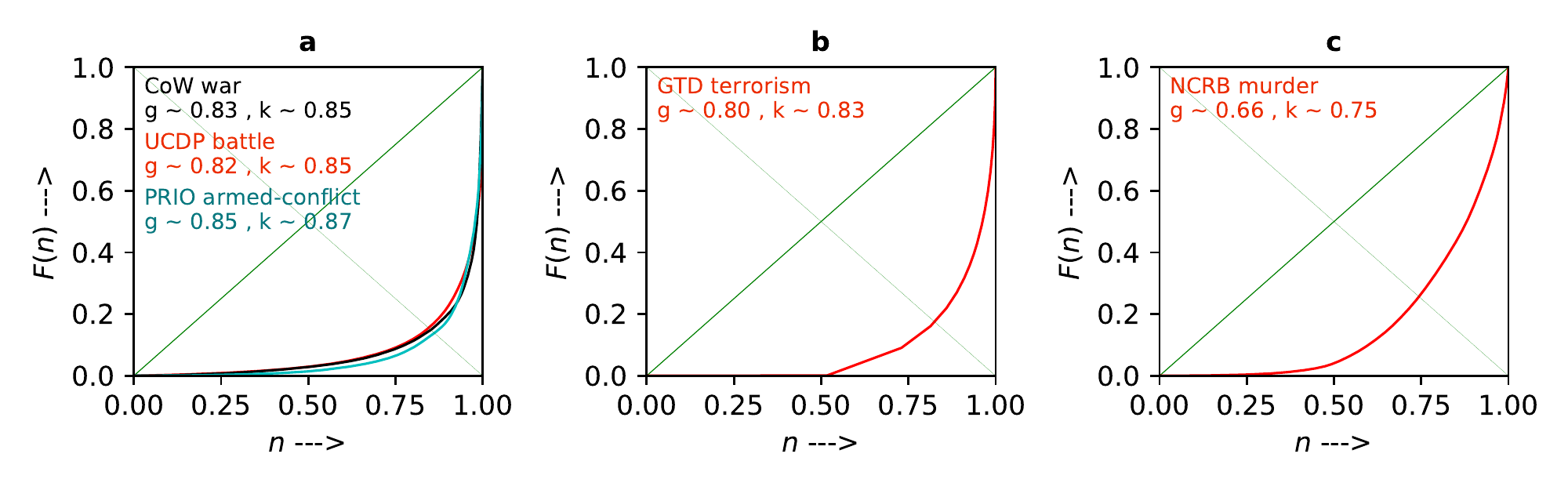}
\caption{Man-made conflicts: Lorenz curve plots for human deaths against fraction of conflicts. (a) War, Battle \& Armed-conflicts: cumulative fraction of deaths place vs. fraction of conflicts according to Correlates of War (CoW) database during 1816-2007, Peace Research Institute Oslo (PRIO) dataset during  1946-2008 and Uppsala Conflict Data Program (UCDP) database for period 1989-2017. In case of PRIO armed conflict dataset (version 3.1) and UCDP battle-death dataset (version 18.0), we used the `best estimate data' to calculate inequality indices (among `best estimate', `low estimate' and `high estimate' fatality columns) (b) Violent Crime: cumulative fraction of people murdered vs. fraction of murder attempts during 1967-2016 based on National Crime Records Bureau of India (NCRB) database, (c) Terrorism: cumulative fraction of victim counts vs. fraction of terror attacks during 1970-2017, as per Global Terrorism Database (GTD).}
\end{figure}

\begin{table}[ht]
    \caption{ Estimated inequality (in death counts) index values for man-made conflicts (see, Table I and Fig. 2) }
    \label{tab:table4}
    \begin{tabular}{|l|c|c|} 
    \hline
      \textbf{Type of conflicts} & \textbf{$g$-index} & \textbf{$k$-index} \\
      \hline
		war & $0.83\pm$0.02 & $0.85\pm$0.02  \\
		\hline     
		battle & $0.82\pm$0.02 & $0.85\pm$0.02 \\
		\hline
      	  armed-conflict & $0.85\pm$0.02 & $0.87\pm$0.02  \\
      \hline
          terrorism & $0.80\pm$0.03 & $0.83\pm$0.02  \\
      \hline
       murder & $0.66\pm$0.02 & $0.75\pm$0.02 \\
		\hline	   
    \end{tabular}
\end{table}

 \subsection{For natural disasters}
 
 Analyzing the data for human death distribution in various natural disasters, as obtained from the datasets mentioned in Table I, we find the Lorenz curves (see, Fig. 3) to be even more steeply curved and the corresponding $g \simeq 0.95 \simeq k$ values are even higher (see, Table III).
 
\begin{figure}[H]
\centering
\includegraphics[width=1.0\textwidth] {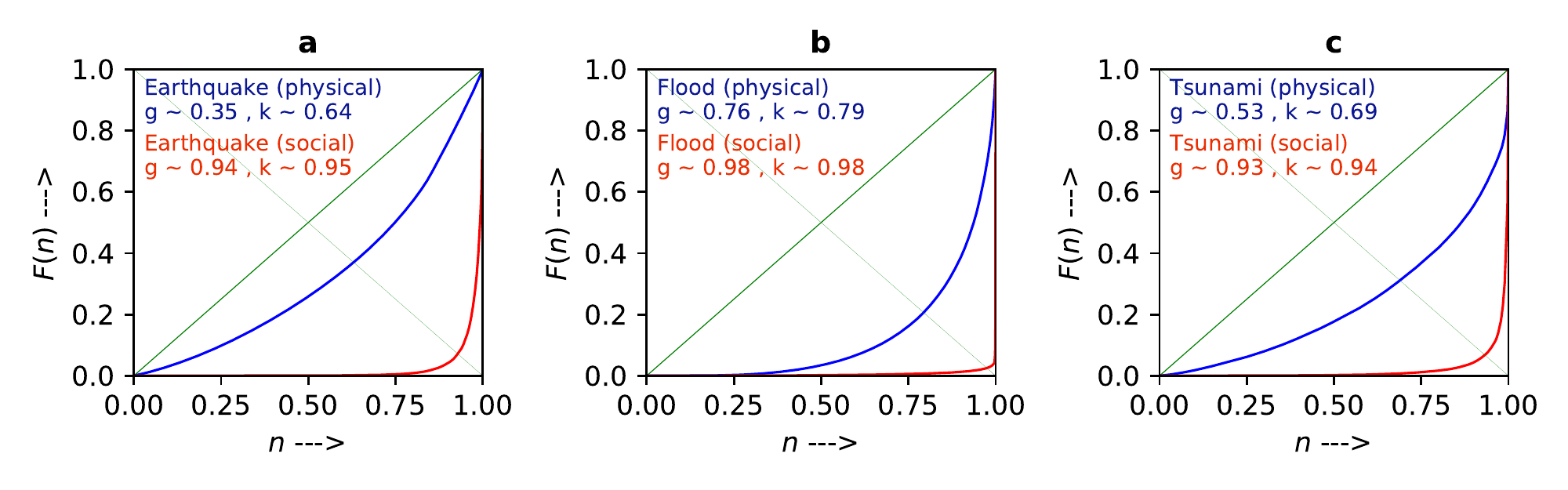} 
\caption{ Natural disasters: Lorenz curve plot of cumulative fraction of damages $F$($n$) for social loss or human deaths (in red) as well as for physical damages (brown), against the fraction ($n$) of natural disasters. (a) Earthquake: cumulative fraction of human deaths during 1000-2018(July) from NCEI-NOAA database and Richter scale magnitudes between 2013-2018(July) according to USGS database against $n$ occurences of earthquake. (b) Flood: cumulative fraction of human deaths and respective areas affected (sq-km) between 1900-2018(July) according to EMDAT database. (c) Tsunami: cumulative fraction of human deaths and corresponding maximum water heights(m) against $n$ occurences of tsunami between 1000-2018(July) according to NCEI-NOAA database.}
\end{figure}
 
  \begin{table}[ht]
    \caption{  Estimated inequality (in death counts) index values for natural disasters (see, Table I and Fig. 3) }
    \label{tab:table4}
    \begin{tabular}{|l|c|c|} 
    \hline
      \textbf{Type of disasters} & \textbf{$g$-index} & \textbf{$k$-index}\\
      \hline
		  earthquake & $0.94\pm$0.02  & $0.95\pm$0.02\\
		\hline     
		 flood & $0.98\pm$0.02 & $0.98\pm$0.02\\
		\hline
      	   tsunami & $0.93\pm$0.02 & $0.94\pm$0.02\\
    
		\hline	   
    \end{tabular}
\end{table}

\begin{table}[H]
  \begin{center}
    \caption{ Inequality (in terms of physical damages) index values of physical damages caused by natural disasters }
    \label{tab:table6}
    \begin{tabular}{|c|c|c|c|c|c|c|c|c|c|c|} 
    \hline    
    
      \textbf{Data source} & \textbf{\makecell{Data\\coverage}} & \textbf{Time period} & \textbf{\makecell{Dataset\\ length }} &  
      \multicolumn{4}{c|} {\textbf{ Damage measures }} &  \textbf{\makecell{Measured\\in}} & \textbf{\makecell{$g$-index\\value}} & \textbf{\makecell{$k$-index\\value}}  \\ \cline{5-8}  & & & &  Min. & Max. & Avg. &  Total & & & \\    
       
       \hline
      \makecell{\textbf{earthquake}\\ \makecell{USGS\\data \cite{usgs}}} & worldwide & 2013-2018(July) & 519959 &   0.5 & 8.3 & 1.96 & 1022601 & \makecell{Richter\\magnitude}  & 0.35 & 0.64 \\
      
	   \hline
		\makecell{\textbf{flood}\\ \makecell{EMDAT\\data\cite{emdat}}} & worldwide & 1900-2018(July) & 1719 & 
 0 & 2857000 & 87868.28 & 151045575  & \makecell{areas\\ \makecell{affected\\(sq-km)}}  & 0.76 & 0.79 \\
		
			\hline       	
		\makecell{\textbf{tsunami}\\ \makecell{NCEI-NOAA\\data \cite{ngdctsunami}}} & worldwide & 1000-2018(July) & 374 &  1 & 524.26 & 10.16 & 3803 & \makecell{maximum\\ \makecell{water\\height (m)}} & 0.53 & 0.69 \\

	   \hline
	   
    \end{tabular}
  \end{center}
\end{table}

However, when the damages are measured in physical quantities (in Richter magnitude for earthquakes, areas affected in floods and water level heights in case of tsunamis) both of the Gini and Kolkata indices assume much lower values (see Fig. 3 and Table IV).

\section{Discussion \& Summary}

\begin{figure}[H]
\centering
\includegraphics[width=0.45\textwidth] {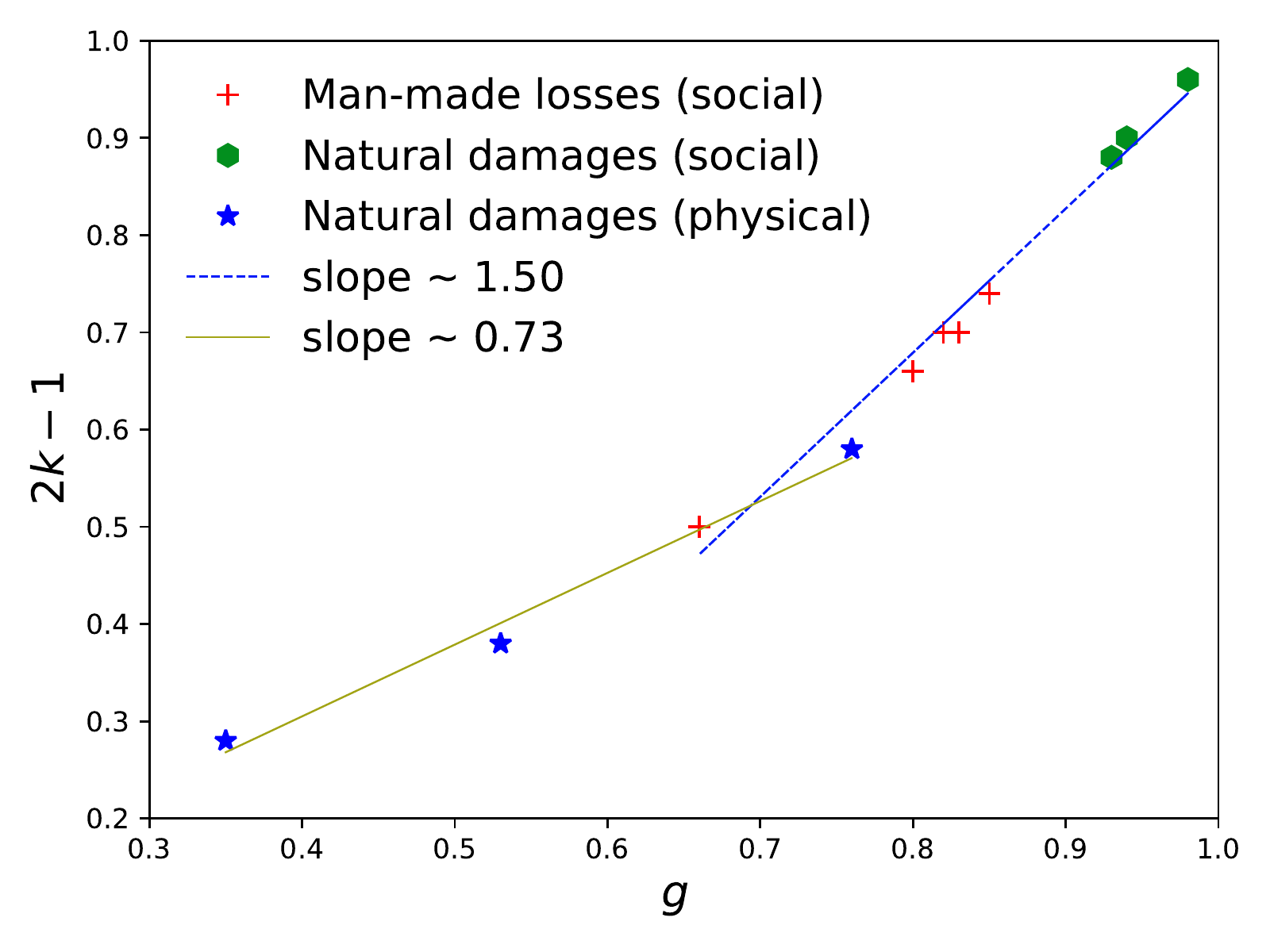} 
\caption{ Plot of the vertical distance between equality line and Lorenz curve ($2k-1$) against Gini index from all the analysed data discussed here (Table II-IV). Both $g$ and $2k-1$ have identical range [0,1]. They seem to emerge two different slopes for the fitting straight lines: about $0.73$ (for lower $g$ range) and $1.5$ (for higher $g$ range). For both the cases, the values of $2k-1$ are seen to be less than the corresponding values of $g$.  }
\end{figure}


Assuming the Lorenz curve to be given by $y = x^\alpha$ for a positive real number $\alpha$, passing through (0,0) and (1,1) points, we get $g$ = $[ \frac{1}{2} - \int_0^1 x^\alpha\;\mathrm{d}x ]/({\frac{1}{2}}) = 1 - \frac{2}{(\alpha+1)}$ and self-consistent equation $k^\alpha = 1-k$ (in Fig. 1, $y$ $\equiv$ $F$ and $x$ $\equiv$ $n$). This gives $g = 0$ and $k = \frac{1}{2}$ for the equality line ($\alpha = 1$) and $g = k = 1$ for extreme inequality ($\alpha\to\infty$). In the social conflict cases studied here, such high values of inequality-index values suggest $\alpha \simeq 10$, giving $g \simeq 0.82$ and $k \simeq 0.84$ (when both $g$ and $k$ are high and close enough in magnitude, as observed). Indeed, even for the so called Pareto 80/20 law (80\% wealth possessed by 20\% of the population \cite{lipovetsky2009pareto}) suggesting $k=0.8$, one gets $\alpha \simeq 7$ ($[0.8]^\alpha = 0.2$, giving $\alpha \simeq 7$) and hence $g$ = $ 1 - \frac{1}{4} \simeq 0.75 $, which also compares well with our earlier observation \cite{ghosh2014inequality} for citation distribution in academic institutions ($g \simeq 0.75$ and $k \simeq 0.8$). As mentioned before, such high values of $g$ and $k$ are not usually seen for income or wealth distributions in societies (economic inequalities; see e.g., \cite{ghosh2014inequality}) though, of late, these values are rapidly increasing  \cite{piketty2014capital}. Human death inequality index values for natural disasters ($g \simeq 0.95 \simeq k$) are even higher. Of course we found that when the social counts (of deaths) are replaced by physical quantities measuring damages in natural disasters (e.g., Richter magnitude in case of earthquakes, amount of areas affected in case of floods or maximum water level heights in case of tsunamis etc.) the inequality index values ($g \leq 0.76$ and $k \leq 0.79$; see Table IV) become somewhat lower. Fig. 4 shows the variations of the quantity $2k-1$ and $g$ for all the cases analysed use (see Table II-IV). Both $2k-1$ and $g$ have identical ranges [0,1]. These seems to be two distinct slopes of the $2k-1$ vs $g$ curves: about 0.73 (for lower $g$ range) and about 1.50 (for higher $g$ range). It may be mentioned that the mapping of Lorenz curve as the quadrant of an unit circle, gives \cite{chatterjee2017socio} $(2k-1)/g \simeq 0.73$, which compares well with the observed initial slope value in Fig. 4. It may be interesting to note that a recent data analysis study on the relationship between Gini index ($g$) and Kolkata index ($k$) has been reported \cite{inskii2019bitcoin} in the context of economic inequality in the bitcoin market. 

In summary, we find deaths from social conflicts are quite inequally distributed. From Table II, we find death inequality measures ($g \simeq 0.82$ and $k \simeq 0.84$) for man-made social conflicts, while from Table III those for natural disasters are $g \simeq 0.95$ and $k \simeq 0.96$. As already mentioned earlier, the citations of the papers produced by various universities or institutes of the world, can be viewed as the wealth created by the respective institutions and these wealths are also found to be quite unequally distributed across the contributions or papers. Our earlier study in \cite{ghosh2014inequality} (see Table V in the Appendix) suggested high level of inequality across the institutions: $g \simeq 0.75$, $k \simeq 0.80$. As the institutions encourage competitiveness (and do not always take care for those who may fall behind or become unsuccessful), the social inequality is very high. Such high values of inequality indices had however not been seen in socio-economic systems earlier ($g \le 0.4$, $k \le 0.6$ \cite{ghosh2014inequality}). These may be due the various welfare measures usually taken by various governments. Also, some entropic considerations suggest \cite{venkatasubramanian2017much} that some amount of inequality may not be unfair in capitalistic societies. Recently the economic inequality has started growing rapidly (since 1950's; see e.g., \cite{piketty2014capital}\cite{factrank}) and index values for both $g$ and $k$ are increasing. This may be because of the fast disappearance of social welfare measures across the world. Our study here indicates a `natural' tendency towards extreme inequality ($g \simeq 0.90 \simeq k$) in our societies, unless some appropriate welfare measures are taken. 
 
\vskip 1 cm
\noindent\textbf{Acknowledgement:} We are thankful to Arnab Chatterjee, Asim Ghosh, Manipushpak Mitra, Sudip Mukherjee for many useful inputs and discussions. AS acknowledges kind hospitality at Saha Institute of Nuclear Physics, Kolkata and BKC acknowledges JC Bose Fellowship (DST) for support.

\section*{APPENDIX}
\begin{table}[H]
  \begin{center}
    \caption{Values of the inequality indices ($g$ and $k$) for some of the academic institutions (from \cite{ghosh2014inequality}, see also \cite{chatterjee2017socio}; source-data taken from the Web of Science)}
    \label{tab:table7}
    \begin{tabular}{|c|c|c|c|c|c|c|c|} \hline    
    
    \textbf{Inst./Univ.} & \textbf{Year}  & 
    \multicolumn{2}{ c|} {\textbf{Index values for}} & \textbf{Inst./Univ.} & \textbf{Year} &  \multicolumn{2}{ c|} {\textbf{Index values for}} 
\\ \cline{3-4} \cline{7-8}   & &    \makecell{Gini\\($g$)} & \makecell{Kolkata\\($k$)}     &&& \makecell{Gini\\($g$)} & \makecell{Kolkata\\($k$)} \\
      
    \hline
      \multirow{4}{*}{Cambridge} &1980 &  0.74 & 0.78 & \multirow{4}{*}{BHU} &1980  & 0.68 & 0.76 \\ &1990 &  0.74 & 0.78 & &1990 &  0.71 & 0.77 \\ &2000 &  0.71 & 0.77 & &  2000 &  0.64 & 0.74\\ &2010  & 0.70 & 0.76  &  &2010 &  0.63 & 0.73 \\  
	  \hline 

      \multirow{4}{*}{Harvard} &1980 &  0.73 & 0.78 & \multirow{4}{*}{Calcutta} &1980 &  0.74 & 0.78 \\ &1990 &  0.73 & 0.78 & &1990 &  0.64 & 0.74\\ &2000 &  0.71 & 0.77 & &2000 &  0.68 & 0.74\\ &2010  & 0.69 & 0.76 & &2010 &  0.61 & 0.73   \\
	  \hline 
	   
      \multirow{4}{*}{MIT} &1980 &  0.76 & 0.79 &  \multirow{4}{*}{Delhi} &1980 &  0.67 & 0.75\\ &1990 &  0.73 & 0.78 & &1990 &  0.68 & 0.76\\ &2000 &  0.74 & 0.78 & &2000 &  0.68 & 0.76\\ &2010 &  0.69 & 0.76  & &2010 &  0.66 & 0.74\\
	  \hline 
	  
      \multirow{4}{*}{Oxford} &1980  & 0.70 & 0.77 & \multirow{4}{*}{IISC} &1980 &  0.73 & 0.78\\ &1990  & 0.73 & 0.78 & &1990 &  0.70 & 0.76 \\ &2000 &  0.72 & 0.77 & &2000 &  0.67 & 0.75\\ &2010 &  0.71 & 0.76  & &2010 &  0.62 & 0.73\\
	  \hline 
	  
      \multirow{4}{*}{Stanford} &1980 & 0.74 & 0.78  & \multirow{4}{*}{Madras} &1980 &  0.69 & 0.76 \\ &1990  & 0.70 & 0.76 & &1990  & 0.68 & 0.76 \\ &2000  & 0.73 & 0.80 & &2000  & 0.64 & 0.73\\ &2010  & 0.70 & 0.76 & &2010  & 0.78 & 0.79  \\
	  \hline 

\multirow{4}{*}{Stockholm} &1980 & 0.70 & 0.76  & \multirow{4}{*}{SINP} &1980 &  0.72 & 0.74 \\ &1990  & 0.66 & 0.75 & &1990  & 0.66 & 0.73\\ &2000  & 0.69 & 0.76 & &2000  & 0.77 & 0.79\\ &2010  & 0.70 & 0.76 & &2010  & 0.71 & 0.76  \\
	  \hline 

      \multirow{4}{*}{Tokyo} &1980  & 0.69 & 0.76 &  \multirow{4}{*}{TIFR} &1980  & 0.70 & 0.76\\ &1990  & 0.68 & 0.76 & &1990 &  0.73 & 0.77 \\ &2000  &  0.70 & 0.76 & &2000  & 0.74 & 0.77\\ &2010  & 0.70 & 0.76 & &2010 & 0.78 & 0.79 \\

	  \hline 
    \end{tabular}
  \end{center}
\end{table}

\end{document}